# BRIGHTNESS AND COLOR OF THE INTEGRATED STARLIGHT AT CELESTIAL, ECLIPTIC AND GALACTIC POLES

Nawar, S., Tadross, A. L., Mikhail, J.S., Morcos, A.B.

*National Research Institute of Astronomy and Geophysics, Helwan, Cairo, Egypt*

## ABSTRACT

From photoelectric observations of night sky brightness carried out at Abu-Simbel, Asaad et al. (1979) have obtained values of integrated starlight brightness at different Galactic latitudes. These data have been used in the present work to obtain the brightness and color of the integrated starlight at North and South Celestial, Ecliptic and Galactic Poles. The present values of the brightness are expressed in $S_{10}(\lambda)$ units and mag/arcsec$^2$. Our results have been compared with that obtained by other investigators using photometric and star counts techniques. The color indices B-V and B-R have been calculated and also compared with the previous published results.

## 1. INTRODUCTION

Few studies for the brightness and color index of the integrated star light at north and south celestial, ecliptic and galactic poles were given by different investigators using star counts and photometric techniques.

Tanabe and Mori (1990) have obtained blue and red brightness of the integrated starlight in 24 sky regions including 4 polar (North celestial, North ecliptic, North and South galactic), using star counts of the blue and red photographs of the Palomar sky survey atlas during the period 1968-1987. The investigation of Roach and Megil (1961), Sharov & Lipaeva (1973) and Tanabe (1973) were also based on star counts.

From observations accumulated by pioneer 10 imaging photo polarimeters from beyond asteroid belt, (Toller et.al. 1987) have obtained integrated starlight at North and south celestial ecliptic and galactic poles in blue and red colors. Also photometric results have been obtained by Elsässer and Hang (1960), Lille (1968) and Classen (1976). The only results for yellow color were given by Tanabe and Mori (1990).

The (B-R) color index at North and south celestial, ecliptic and galactic poles have been studied by Tanabe (1973) and Tanabe and Mori (1976). Zavarzin (1975) has obtained the (B-R) color Index only at North galactic pole. The (B-V) color index has only been obtained by Tanabe and Mori (1990). In the present work the brightness, (B-V) and (B-R) color indices of the integrated starlight at North and South celestial, ecliptic and galactic poles will be studied using the published data by Asaad et al. (1979).

## 2. WHY THE POLES ?

The integrated starlight at different poles is very important in studies of the light of night sky. The celestial poles, especially the North Pole have been used in a number of studies of airglow line and continuum covariance groups and short and long term variations in the light of night sky.

The ecliptic poles are important regions in studies of zodiacal light. The photometric data at these positions provide information on the scattering properties of interplanetary dust at high ecliptic latitudes and also in measuring the thickness of the zodiacal cloud. Background starlight and zodiacal light have comparable brightness at the ecliptic poles and cannot be easily separated without large errors. For this reason isolation of the zodiacal light at the ecliptic poles in the ground and space observations is possible if the color and distribution of the background starlight in these regions are known.

Background starlight and diffuse galactic light brightness are minima near the galactic poles which have been also observed from the earth and space. Therefore integrated starlight data may be combined with the diffuse galactic light at galactic poles to accurately isolate the cosmic light.

## 3. OBSERVATIONS AND RESULTS

From the photoelectric observations of the night sky brightness carried out at Abu-Simbel in 1975-1977, Asaad et al. (1979) have obtained the distribution of the integrated starlight with galactic latitudes. Their results were expressed in $S_{10}$ (B), $S_{10}$ (V) and $S_{10}$ (R) for blue, yellow and red colors respectively. Where $S_{10}$ ($\lambda$) is the number of stars of $10^{th}$ magnitude per square degree at certain wavelength ($\lambda$). Using these data with simple interpolations, the brightness of the integrated starlight at north and south celestial, ecliptic and galactic poles have been obtained for blue ($\lambda_{eff}$ =4410 Å), yellow ($\lambda_{eff}$ =5500 Å) and red ($\lambda_{eff}$=7900 Å). The results are given in Table (1) in $S_{10}$ ($\lambda$) units. These results have been converted to mag/arcsec$^2$ using the following relation. The results are also given in Table (1).

$$I (mag/arcsec^2) = -2.5 \log (I (S_{10} \lambda)) + 27.78$$

(cf. Nawar et al. 1997, 2007)

The present values of the integrated starlight in yellow are new values after Tanabe and Mori (1990), while the values at far red at $\lambda$=7990 Å are obtained for the first time as there is no previous published data for the integrated starlight at this wavelength.

Table 1: Brightness of the integrated star light obtained at North and South celestial, ecliptic and galactic poles for Blue, Yellow, and Red colors.

| Region | Blue | | Yellow | | Red | |
|---|---|---|---|---|---|---|
| | S10 (B) | mag | S10 (V) | mag | S10 (R) | mag |
| NCp | 37 | 23.86 | 63 | 23.28 | 199 | 22.03 |
| SCp | 41 | 23.74 | 64 | 23.26 | 194 | 22.06 |
| NEp | 31 | 24.05 | 60 | 23.33 | 140 | 22.41 |
| SEp | 38 | 23.83 | 60 | 32.33 | 153 | 22.32 |
| NGp | 15 | 24.84 | 26 | 24.24 | 41 | 23.75 |
| SGp | 15 | 24.84 | 30 | 24.08 | 51 | 23.51 |

## 4. DISCUSSION OF THE RESULTS

It can be seen from Table (1) that the dark part in the sky is at north galactic pole (NGP) and south galactic pole (SGP). Table (1) also indicates that, for blue color there is no difference in the brightness of the integrated starlight at NGP & SGP, while there is slight difference about 0.12 mag/arsec$^2$ between north celestial pole (NCP) and south celestial pole (SCP). The difference increase to 0.22 between north ecliptic pole (NEP) and south ecliptic pole (SEP).

For yellow color the values of the integrated starlight at NCP & SCP and NEP & SEP are almost the same. While there is slight difference about 0.16 mag/arcsec$^2$ between NGP & SGP.

For red color the integrated starlight values at NCP&SCP are almost the same. The difference in the brightness between NEP & SEP is 0.09 mag/arcsec$^2$ and between NGP & SGP is 0.24 mag/arcsec$^2$.

## 5. COMPARISON WITH THE PREVIOUS RESULTS

The present results of the integrated starlight have been compared with that obtained by other investigators for blue, yellow and red colors. The published results by different investigators are expressed in $S_{10\odot}$ units. To held the comparison it is necessary to convert their results from $S_{10\odot}$ to $S_{10}$ (B), $S_{10}$ (V) and $S_{10}$ (R) for blue, yellow and red colors respectively. These have been done using conversion factor for visual brightness units given in Table 2 page 3 from Leinert et al. (1997). The results of other investigators have been also converted to mag/arcsec$^2$ and given in Tables 2 - 4 for blue, yellow and red colors.

It can be seen from Tables 2 & 3, using photometric technique the values of the integrated starlight at the poles expressed in $S_{10}$ (B) units are higher than that obtained by star counts technique. The reason is due to the errors arising when using the two different techniques especially star counts.

The results of Tables 2-4 are drawn in Figs.1- 3. Fig 1 gives for blue color the comparison between our results with that obtained by different investigators using photometric technique. It can be seen from Fig.1 that at NGP the higher value of integrated starlight expressed in mag/arcsec$^2$ obtained by Classen (1976) and the lower value obtained by Lillie (1968). The difference in the brightness between them is 0.28 mag/arcsec$^2$. Our results are similar to that obtained by Toller (1981). The same results have been obtained at NGP except the higher value at Toller (1987).

Table 2: Comparison between the present results of the integrated star light at different poles for blue filter with that obtained by other investigators using photometric technique

| Author | Present work | | Lillie 1968 | | Classen 1976 | | Toller 1981 | | Toller 1987 | |
|---|---|---|---|---|---|---|---|---|---|---|
| Region | S10 (B) | mag | S10 (B) | mag | S10 (B) | mag | S10 (B) | mag | S10 (B) | mag |
| NCp | 37 | 23.86 | 30 | 24.09 | | | 40 | 23.77 | 31 | 24.05 |
| SCp | 41 | 23.74 | | | 30 | 24.09 | 46 | 23.62 | 41 | 23.74 |
| NEp | 31 | 24.05 | 52 | 23.4 | | | 35 | 23.92 | 36 | 23.89 |
| SEp | 38 | 23.83 | | | 48 | 23.58 | 42 | 23.72 | 70 | 23.17 |
| NGp | 15 | 24.84 | 14 | 24.29 | 12 | 25.12 | 15 | 24.84 | 16 | 24.77 |
| SGp | 15 | 24.84 | 18 | 24.64 | 15 | 24.84 | 15 | 24.84 | 14 | 24.91 |

Table 3: Same as Table 2 using star counts technique.

| Author | Present work | | Roach & Megil 1961 | | Tanabe 1973 | | Sharov & Lipeava 1973 | | Toller 1981 | | Tanabe 1990 | |
|---|---|---|---|---|---|---|---|---|---|---|---|---|
| Region | S10 (B) | mag | S10 (B) | mag | S10 (B) | mag | S10 (B) | mag | S10 (B) | mag | S10 (B) | mag |
| NCp | 37 | 23.86 | 26 | 24.24 | 26 | 24.24 | 20 | 24.52 | 40 | 23.77 | 22 | 24.43 |
| SCp | 41 | 23.74 | 30 | 24.07 | | | 31 | 24.05 | 46 | 23.62 | | |
| NEp | 31 | 24.05 | 28 | 24.16 | 35 | 23.92 | 28 | 23.15 | 35 | 23.92 | 29 | 24.13 |
| SEp | 38 | 23.83 | 27 | 24.2 | | | 21 | 24.16 | 42 | 23.72 | | |
| NGp | 15 | 24.84 | 13 | 25 | 14 | 24.91 | 12 | 25.08 | 15 | 24.84 | 18 | 24.66 |
| SGp | 15 | 24.84 | 15 | 24.83 | 28 | 24.16 | 13 | 25 | 15 | 24.84 | 21 | 24.47 |

Table 4: Same as Table 2 for red color using photometric & star counts techniques

| Author | Present work | | Pioner photometry | | Tanabe 1973 | | Toller 1981 | | Tanabe & Mori 1990 | |
|---|---|---|---|---|---|---|---|---|---|---|
| Region | S10 (R) | mag | S10 (R) | mag | S10 (R) | mag | S10 (R) | mag | S10 (R) | mag |
| NCp | 199 | 22.03 | 109 | 22.62 | 108 | 22.7 | 124 | 22.5 | 67 | 23.21 |
| SCp | 194 | 22.06 | 133 | 22.47 | | | 148 | 22.35 | | |
| NEp | 140 | 22.41 | 116 | 22.62 | 108 | 22.7 | 108 | 22.67 | 70 | 23.16 |
| SEp | 153 | 22.32 | 177 | 22.16 | | | 134 | 22.46 | | |
| NGp | 41 | 23.79 | 44 | 23.67 | 48 | 23.72 | 36 | 23.84 | 60 | 23.33 |
| SGp | 51 | 23.51 | 51 | 23.51 | 112 | 22.65 | 37 | 23.86 | 67 | 23.21 |

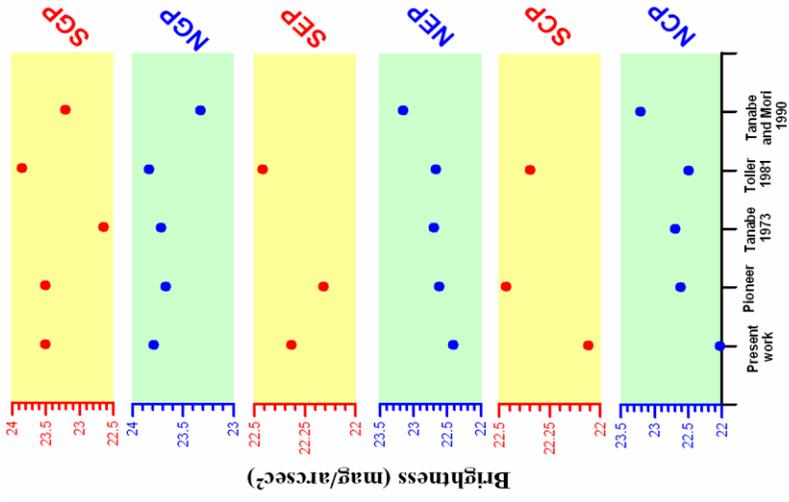

Fig 3. Same as Fig 1 for red color using star counts and photometric tichniques.

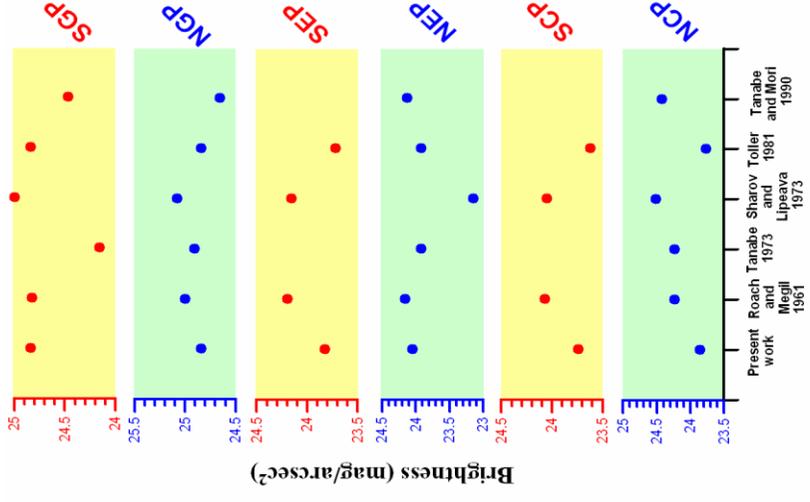

Fig 2. Same as Fig 1 for blue color using star counts tichnique.

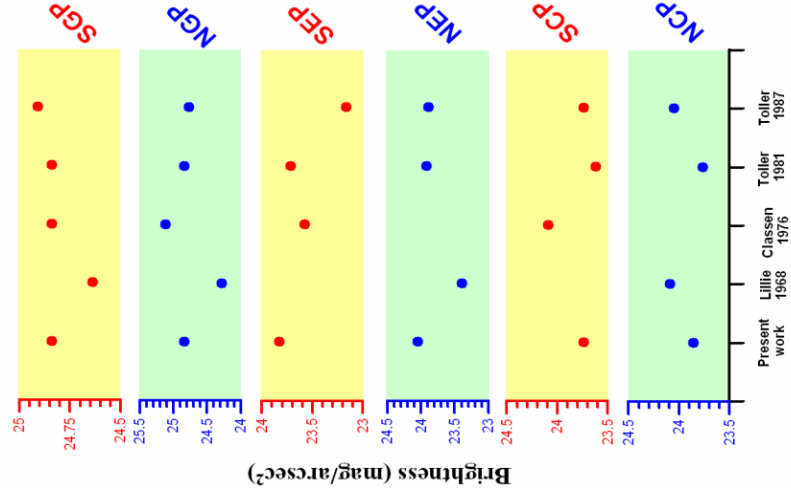

Fig 1. Brightness of the intgrated star light obtained by different investigators for blue color using photometric tichnique.

At NEP & SEP our results of integrated starlight expressed in mag/arcsec$^2$ is higher than that obtained by other investigators. While the lower values at NEP and SEP can be seen in the Fig. 1 at Lillie (1968) and Toller (1987) respectively.

At NCP slight difference has been obtained between our values and that of other investigators. This difference is no more than 0.2$^m$. While at SCP the present values are almost the same with that obtained by other investigators, but it is lower than that of Classen (1976).

Fig. 2 is similar to Fig.1 but for star counts. It can be seen from this figure that, at NGP&NEP&SCP and NCP the present values of the integrated starlight expressed in mag/arcsec$^2$ are almost the same with that obtained by Toller (1981) and lower than that obtained by other investigators.

As shown in Fig. 3, at NCP and NEP, the present B-R values of the integrated star light are the lower one, while that of Tanabe and Mori (1990) are the higher values. And vice versa at NCP, our value is the higher one, while that of Tanabe and Mori (1990) is the lowest value. The discrepancies in the previous results may be due to the fact that the present results, which are from ground-based observations, while that obtained by Tanabe and Mori (1990) were taken from outer space.

## 6. COLOR INDEX

From the results of Tables 2-4 the B-V and B-R of the integrated starlight have been obtained using the following two relations:

$$B-V = -2.5 \log (I_B/I_V)$$

$$B-R = -2.5 \log (I_B/I_R)$$

The results are given in Table 5. The present results of B-V have only compared with that obtained by Tanabe and Mori (1990). The reason is due to there are no published results in V color for the

BRIGHTNESS AND COLOR OF THE INTEGRATED STARLIGHT AT
CELESTIAL, ECLIPTIC AND GALACTIC POLESintegrated starlight brightness at different poles except Tanabe and Mori (1990). It can be seen from Table 5 that except NEP the present values of B-V at NCP&NGP and SGP are lower than that obtained by Tanabe and Mori (1990). The reason may be due to the difference in the used technique. In the present work photoelectric technique is used, while Tanabe and Mori (1990) used star counts technique.

It can be seen also from Table 5 that NCP and SCP is redder than other ecliptic and galactic poles. Our B-R values of the integrated starlight are higher than that obtained by other investigators. This means that the color of the integrated starlight obtained in the present work is redder than that of other investigators. The reason is due to the big difference between our red wavelength (far red λ=7900 Å) and that of other investigators.

Table 5a: Comparison of (B-V) obtained in the present work with that obtained by other investigators.

| Author | Present W | Tanabe 1990 |
|---|---|---|
| NCp | 0.58 | 0.76 |
| SCp | 0.48 | |
| NEp | 0.72 | 0.61 |
| SEp | 0.5 | |
| NGp | 0.6 | 0.79 |
| SGp | 0.76 | 0.79 |

Table 5b: Same as Table 5a for (B-R).

| Author | Present W | Raoch & Megil 1961 | Tanabe 1973 | Toller 1981 | Pionner 10 (1987) | Tanabe & Mori 1990 |
|---|---|---|---|---|---|---|
| NCp | 1.83 | 1.3 | 1.54 | 1.27 | 1.36 | 1.22 |
| SCp | 1.68 | 1.36 | | 1.27 | 1.27 | |
| NEp | 1.64 | 1.32 | 1.22 | 1.25 | 1.27 | 0.97 |
| SEp | 1.51 | 1.37 | | 1.26 | 1.01 | |
| NGp | 1.09 | 1.29 | 1.19 | 0.95 | 1.1 | 1.33 |
| SGp | 1.33 | 1.27 | 1.51 | 0.98 | 1.4 | 1.26 |

## CONCLUSION

Brightness and B-V & B-R color indices of the integrated starlight at north and south celestial, ecliptic and galactic poles have been obtained for blue, yellow and red colors. The results of the brightness are expressed in $S_{10}$ (B) & $S_{10}$ (V) and $S_{10}$ (R) units and mag/arcsec$^2$. The present results have been compared with that obtained by other investigators after converting their values from $S_{10}$ ☉ to $S_{10}$ (B), $S_{10}$ (V) & $S_{10}$ (R), units and mag/arcsec$^2$. The present values of B-V and B-R have been also compared with the previous results of other investigators.

## REFERENCES

Asaad, A. S. and Nawar, S.: 1979, Hel. Obs. Bull. No. **200.**

Classen, C.:1976, Eine Blae-Photometric der Milchstrasse und des Zodiakallichts, Inaugural- Dissertation zur Erlangung des Doktorgrades, Bonn.

Elsasser, H., Haug, U.: 1960, Z.Astrophys. **50**, 121.

Leinert, C. H. et al.(14 coauthors.)1997, yCat, **41270001L.**

Lillie, C. F.:1968, An Empirical Determination of the Interstellar Radiation Field, .Ph.D. Thesis University of Wisconsin.

Nawar, S. and Morcos, A.B.; 1997, AS & SS. **253**, 1.

Nawar, S.; Morcos, A. B.; Mikhail, J. S.; 2007, NewA. 12, 562.

Roach, F. E. and Megill, I.R.: 1961, Astrophys. J. **133**,228.

Sharov, A. S. and Lipaeva, N. A.:1973, Soviet Asron. **17**, 69.

Tanabe, H.:1973, World data center $C_2$ (airglow), Tokyo Astron. Obs. Mitaka, Japan, **45**.

Tanabe, H. and Mori, K.: 1990, IAUS, 1**39**, 103.

Toller, G. N.:1981, Ph.D. Dissertation, State University of N.Y.

Toller, G. N.: 1987, A&A, **88**, 24.

Zavarzin, Yu. M.: 1975, Astron. Tsrik, **874**, 6.